\documentclass[aps,onecolumn]{revtex4}
\usepackage{amsfonts,amssymb,amsmath,bm}
\usepackage[dvips]{graphicx}
\usepackage{color}

\begin{document}

\title{Long-range interactions between membrane inclusions: Electric field induced giant amplification of the pairwise potential}

\author{E. S. Pikina$^{1,2}$, , A.R. Muratov$^2$, E. I. Kats$^1$, V. V. Lebedev$^{1,3}$.}

\affiliation{$^1$ Landau Institute for Theoretical Physics, RAS, \\
142432, Chernogolovka, Moscow region, Russia, \\
$^2$  Oil and Gas Research Institute, RAS, 119917, Gubkina 3, Moscow, Russia,\\
$^3$ NRU Higher School of Economics, \\
101000, Myasnitskaya 20, Moscow, Russia.}

\begin{abstract}

The aim of this work is to revisit the phenomenological theory of the interaction between membrane inclusions, mediated by the membrane fluctuations. We consider the case where the inclusions are separated by distances larger than their characteristic size. Within our macroscopic approach a physical nature of such inclusions is not essential, however
we have always in mind two prototypes of such inclusions: proteins and RNA macromolecules. Because the interaction is driven by the membrane fluctuations, and the coupling between inclusions and the membrane, it is possible to change the interaction potential by external actions affecting these factors. As an example of such external action we consider an electric field. Under external electric field (both dc or ac), we propose a new coupling mechanism between inclusions possessing dipole moments (as it is the case for most protein macromolecules) and the membrane. We found, quite unexpected and presumably for the first time, that the new coupling mechanism yields to giant enhancement of the pairwise potential of the inclusions.
This result opens up a way to handle purposefully the interaction energy, and as well to test of the theory set forth in our article.

\end{abstract}


\maketitle

\section{Introduction}
\label{sec:intro}

It is a  great honor, painted by sadness, for us to contribute to this special memorial issue of Annals of Physics devoted to I.E.Dzyaloshinskii, one of the greatest physicist of the twentieth century. All of us had rewarding pleasure to discuss with Dzyaloshinskii various scientific problems and two of us (E.K., and A.M.) were his students and coauthors. We do believe that our work (fluctuation induced interactions) fits the topics selected for this special issue, and as well as Dzyaloshinskii own pioneering and groundbreaking contributions to the theory of Van der Waals forces.

The interplay between various membrane inclusions and the membrane shape deformations has been subject to recurrent focus during last three decades (see, e.g., \cite{GB93,PL96}). Generally, a lot of mechanisms of coupling of membrane inclusions to various degrees of freedom of the membrane can be forecast: curvature, thickness, membrane lipid composition, tilt, and so on. However having in mind biologically relevant inclusions such as proteins, there is currently no theoretical consensus
(\cite{GB93,GG96,PL96,FD97,NE97,DF99,DF02,KG99,YD12,HW01,GW21,MN08}) about which physical mechanisms are dominating these interactions. Since this subject continues to be of considerable interest in what follows we review the different classes of membrane mediated interactions related to different types the inclusion - membrane coupling.

Two large classes of macromolecules, namely the proteins and RNA molecules have been identified, as particularly biologically important membrane
inclusions. In what follows we will use interchangeably the both terms. Inclusions when discussing generic features of our
macroscopic theory, and proteins, speaking about biological consequences of the theory. Forces induced by the membrane fluctuations are interesting in its own right phenomenon covering a huge diversity of topics ranging from down to the earth solid state physics and through cosmology or astrophysics (see many fascinating examples in the review paper \cite{KG99}).
Inclusions embedded into a membrane are ideal objects to study the fluctuations mediated forces between those inclusions. To revisit the phenomenological theory of the interaction between membrane inclusions is one motivation for our work. The second motivation is related to essential biological realization of the proteins included into the membrane. The matter is that the membrane proteins exhibit at some conditions a tendency of phase separation. As a result, the domains of the dense protein phase with liquid, gel or solid structures are formed. Protein clustering in the cell membranes is vital for its biological functions (see, e.g., \cite{MN08,FK09,GW21,KC14,ML15,MC18} and references therein). The protein domains emerge either spontaneously or under an influence of specific driving forces \cite{FK09}. To support the domains formation mutual attracting interaction forces between the macromolecules have to overcome the entropic tendency to homogenize the system.

Although in living matter nonequilibrium processes play a crucial role, often one may select a relatively small subsystem (e.g., a membrane) which are sufficiently close to thermodynamic equilibrium. Then the subsystem can be characterized by temperature, and its fluctuations can be analyzed
within the Gibbs distribution. Certainly the equilibrium approach can be applied to the inclusions in artificial lipid membranes.
In such a case classical theories relying on concepts of equilibrium statistical mechanics,
can be used to describe also some features of non-equilibrium phenomena. Then the main driving effect
for phase separation or aggregation is direct or indirect intermolecular interactions \cite{LI01}.
The strength of the intermolecular interactions may be controlled by several means.

Here, we focus on physical mechanisms underlying biological processes driven by protein aggregation or phase separation,
with a particular attention to the role played by the membrane-mediated intermolecular interactions. Although the membrane proteins in a liquid-like membrane are free to diffuse in the cell membrane the inter-protein interactions via lipid bilayer can essentially influence their organization and thus have an impact on many aspects of their activity.
Hence, coupling of the proteins to the host membrane, as well as the resulting protein-protein interactions, are fundamentally important topics in biophysics.

To name a few we mention the so-called protein distillation, by its result analogous
to the well known classical distillation (i.e., the process of separating the substances from a liquid mixture by using boiling and condensation). For proteins in cell membranes the distillation (it may be a partial separation that increases the concentration of selected components) occurs not by boiling but due
to specific active means developed by molecular motors which transport membrane proteins  towards appropriate destinations. However, irrespectively to biologically active mechanisms involved into the distillation process (see, e.g., \cite{ZV21}) a study of the physical (equilibrium or ``passive'') mechanisms of interactions between proteins is the mandatory first step.
There is also another motivation to investigate the driven interaction forces for the protein ordering. The fact is that the protein crystallization allows researchers to study structural characteristics of the proteins \cite{CS08,MG14,Gordeliy}. And the first step to analyze the onset of the protein crystallization is to find their interaction characteristics in the initially dilute limit.

In the previous works the membrane surface tension was not taken into account, assuming tacitly that for a large lateral size membrane, formed spontaneously by lipid self-organization, the membrane surface tension effectively vanishes \cite{SA94}. However, the membranes in the biological cells are not an isolated infinite lateral size membrane. They are mechano-biological units that encompass the membrane itself, its interacting proteins, and the complex underlying cytoskeleton. Recently, attention has been directed to the membrane tension, which has been linked to diverse cellular processes (see e.g., the paper \cite{SD20}, entitled ``Pay attention to membrane tension''). Even in more simple model lipid membranes, their surface tension could be small but non-zero due to external fields or boundary conditions. The finite surface tension can change qualitatively the membrane mediated interaction energy.

As we said already above one can envisage many physical mechanisms providing the coupling of an inclusion to the membrane. They can be classified in accordance with the symmetry of the inclusions. To be specific in this work we restrict ourselves to the following classes:
\begin{itemize}
\item
(i)
Quadratic in the membrane curvature coupling of the up-down symmetric and in-plane isotropic inclusions;
\item
(ii)
Quadratic in the membrane curvature coupling of the up-down symmetric and in-plane anisotropic inclusions;
\item
(iii)
Linear over the membrane curvature coupling of the up-down asymmetric inclusions (both, in-plane isotropic, or
possessing an in-plane anisotropy);
\item
(iv)
Electric field induced coupling of bearing out-of-plane dipole moment inclusions.
\end{itemize}
Surprisingly for us we did not find publications discussing the last case in the literature. Motivated by this fact we revisit a phenomenological theory of membrane mediated interactions.

For completeness and convenience of potential readers of the paper, we describe shortly also the known (for stressless membranes) results, derived by our method. In the next section \ref{sec:fluct} we formulate our approach and present the obtained expressions for the interaction energy. Namely: In Subsection \ref{subsec:i} we find the pairwise interaction energy mediated by a stressed membrane shape fluctuations between the up-down symmetric and in-plane isotropic inclusions.
The main contribution into the interaction potential comes from the quadratic in membrane curvature coupling between inclusions and the membrane.
Similar (quadratic over membrane curvature) coupling mechanism is considered in Subsection \ref{subsec:ii}
for in-plane anisotropic (quadrupolar) inclusions.
Subsection \ref{subsec:iii} is devoted to the class (iii). The dominating coupling mechanism for the up-down asymmetric
inclusions is linear over curvature. We presented in this subsection the expressions for the pair-wise potential for isotropic in-plane,
and quadrupolar in-plane inclusions embedded into the stressed membrane.
Section \ref{sec:electric} contains the main new message of our work. We show that the external electric field (both dc or ac) strongly increases the membrane mediated interaction provided the inclusion possess an electric dipole moment (as it is the case for the protein macromolecules).
We present some qualitative dimensional estimations for the phenomenological coupling constants entering our expressions
in Section \ref{sec:estim}. Within the classes (i) - (iii) we recover the known results for tensionless ($\sigma =0$) membranes \cite{GB93,GG96,PL96,FD97,NE97,DF99,KG99,YD12,HW01,GW21}). We close with a conclusion discussion in Section \ref{sec:con}.
In the absence of pedagogical textbooks describing the method of calculations (apparently reinvented a few times in various forms), we present the full technical details necessary to perform the actual calculations in Appendix \ref{sec:pair}.

\section{An inter-particle interaction induced by membrane fluctuations}
\label{sec:fluct}

We consider the membrane at scales larger than their thickness. Then the membrane can be treated as a two-dimensional sheet of variable shape. The fluctuations of the membrane shape are controlled by Helfrich energy
\begin{equation}
{\cal F}_H=\int dS\,\left[
\sigma +\frac{\kappa}{2}\left(\frac{1}{R_1}+\frac{1}{R_2}\right)^2
+\frac{\bar\kappa}{R_1R_2}\right],
\label{Helfrich}
\end{equation}
where $dS$ is the element of the membrane area, $R_1$ and $R_2$ are the membrane principal curvature radii, $\sigma$ is the surface tension and $\kappa,\bar\kappa$ are bending and Gaussian rigidity moduli, termed traditionally as Helfrich moduli. We assume that the membrane has up-down symmetry, that is why the term proportional to the mean curvature $R_1^{-1}+R_2^{-1}$ is absent in the expression (\ref{Helfrich}).

The energy (\ref{Helfrich}) was introduced in the original paper \cite{HE73}, see also its textbook versions \cite{KL93,SA94,GP93,NT00,CL00}. In the lipid membranes the surface tension is usually small, that is the length $\sqrt{\kappa/\sigma}$ is much larger than the membrane thickness.
The thermodynamic stability of flat membranes obviously requires the condition $\kappa > 0$. At the same time stability against formation of separate vesicles from a membrane and against the growth of saddles of mean zero curvature means that the Gaussian rigidity should be restricted by the following inequalities $0 > {\bar \kappa } > -2\kappa $.

For relatively small membrane shape fluctuations, the membrane is approximately flat, we choose the $Z$-axis to be perpendicular in average to the membrane. The condition restricts lateral size scales, which should be smaller than the membrane persistence length \cite{SA94,GP93} $\xi \sim h_0 \exp (4\pi \kappa /3 k_B T)$, where $h_0$ is the equilibrium membrane thickness. Since for real lipid membranes $\kappa$ is essentially greater than $k_B T$ the persistence length is larger than all the characteristic scales we are interested in. There is also another length scale $l_{\sigma }=(\kappa /\sigma )^{1/2}$, related to the surface tension $\sigma$. At the scales $r\ll\sqrt{\kappa/\sigma}$ the surface tension $\sigma$ weakly influences the membrane properties.

Fluctuations of the membrane shape are described by its displacement $u(x,y)$ in the $Z$-direction. In the second order in $u$ the energy (\ref{Helfrich}) becomes
\begin{equation}
{\cal F}_u=\int dx\, dy\, \left\{
\frac{\kappa}{2} (\nabla^2 u)^2
+\frac{\sigma}{2} (\nabla u)^2\right\}.
\label{energy1}
\end{equation}
Here we used the two-dimensional differential operator $\nabla$: $(\nabla u)^2=(\partial_x u)^2 + (\partial_y u)^2$, $\nabla^2 =\partial_x^2+\partial_y^2$ and so further. In the approximation  (\ref{energy1}) the field $u$ possesses Gaussian statistics, and is completely characterized by its pair correlation function. The explicit expression for the pair correlation function is presented in Appendix \ref{sec:pair}.

Note that there is no contribution to the energy (\ref{energy1}) related to the Gaussian curvature term with the modulus $\bar\kappa$ in Eq. (\ref{Helfrich}). This contribution into the membrane energy (\ref{Helfrich}) depends solely on the membrane topology \cite{KL93,SA94,NT00,CL00}, and therefore it is unchanged by the small membrane shape fluctuations. The property is a consequence of Gauss-Bonnet theorem.

The energy of the membrane containing inserted protein molecules can be represented as
\begin{equation*}
{\cal F}={\cal F}_u+{\cal F}_{int},
\end{equation*}
where ${\cal F}_u$ is the energy of the membrane fluctuations and ${\cal F}_{int}$ is the coupling energy of the protein molecules to the membrane. As it is demonstrated in the papers \cite{Mouritsen84,Gordeliy} the coupling energy ${\cal F}_{int}$ is sensitive to the mismatch between the membrane thickness and the length of the hydrophobic part of the protein molecule, so-called hydrophobic mismatch $\epsilon$. There are also other factors determined the interaction energy. They will be discussed in Section \ref{sec:estim}.

We consider the case of a small protein concentration. In this case the interaction energy between the inclusions is determined by the membrane fluctuations at the distances $r \gg a$, where $a$ is a characteristic size of the inclusion along the membrane. Thus we can treat the protein molecules as point-like objects inserted into the membrane, and the interaction energy ${\cal F}_{int}$ is a sum of the terms related to the protein molecules and dependent on their positions. One may consider different contributions to the interaction energy ${\cal F}_{int}$, that are determined by the coupling mechanisms between the protein molecules and the membrane. Below, for the classes (i)-(iv) listed in Section \ref{sec:intro} we derive the pairwise interaction energy for the corresponding coupling mechanisms.

To find the interaction potential between the proteins $U$, mediated by the membrane fluctuations, one should start with the general expression for the free energy
\begin{equation}
\exp\left(-F/T\right)=
\int {\cal D}u\, \exp\left(-{\cal F}/T\right),
\label{interaction}
\end{equation}
where the integral is performed over the membrane fluctuations. In the second order approximation over the interaction energy ${\cal F}_{int}$ the potential $U$ is
\begin{equation}
U=-\frac{1}{2T}
\left\langle {{\cal F}}_{int}^2 \right\rangle,
\label{energy}
\end{equation}
where angular brackets mean averaging over the fluctuations of the membrane. The approximation is justified by smallness of the membrane fluctuations. Starting from the expressions (\ref{energy1}, \ref{interaction}) it is possible to study the membrane mediated interaction between the protein molecules.

\subsection{Quadratic in curvature, up-down symmetric coupling.}
\label{subsec:i}

We analyze different contributions to the coupling energy of the proteins with the membrane. In this Subsection we consider the interaction between the up-down symmetric inclusions, case (i) in Section \ref{sec:intro}. We assume here that the inclusions are isotropic, then the interaction between the membrane and the inclusions has to be isotropic as well. In the main approximation it can be determined by the mean curvature $R_1^{-1}+R_2^{-1}$ or by Gaussian curvature $(R_1 R_2)^{-1}$ (see the membrane energy (\ref{Helfrich})). The curvature should be taken at the point where the inclusion is inserted.

For the up-down symmetric inclusions the interaction energy ${\cal F}_{int}$ is quadratic in $u$. In the linear in $u$ approximation $R_1^{-1}+R_2^{-1}\approx \nabla^2 u$, therefore the interaction term, proportional to $(R_1^{-1}+R_2^{-1})^2$, can be written as 
\begin{equation}
{\cal F}_{Bint}= \sum_j  \,B_j \left[\nabla^2 u(\bm r_j)\right]^2.
\label{inter1}
\end{equation}
Here the summation is performed over positions of the protein molecules, $\bm r_j=(x_j,y_j)$, and $B_j$ are coupling constants. The constants $B_j$ cannot be found within our macroscopic approach, only certain heuristic estimations are possible, see Section \ref{sec:estim}.

Now we proceed to Gaussian curvature. In the main approximation in $u$
\begin{equation*}
\frac{1}{R_1R_2}\to
\frac{1}{2}\epsilon_{ik}\epsilon_{mn}
\partial_i\partial_m u
\partial_k\partial_nu
=\partial_x^2u \partial_y^2 u -(\partial_x \partial_y u)^2,
\end{equation*}
where $\epsilon_{ik}$ is the antisymmetric tensor, $\epsilon_{xy}=-\epsilon_{yx}=1$. Thus in the second order in $u$ we find the following contribution to the interaction energy
\begin{equation}
{\cal F}_{Dint}= \sum_j
\frac{1}{2}\epsilon_{ik}\epsilon_{mn}
\partial_i\partial_m u
\partial_k\partial_nu .
\label{energy4}
\end{equation}
The term (\ref{energy4}) has the structure similar to one of the term (\ref{inter1}).

Using the expression (\ref{energy}), one can derive the interaction energy between two proteins, $1$ and $2$. Performing straightforward calculations with the expressions (\ref{oper},\ref{oper1},\ref{oper2}) we find
\begin{equation*}
\langle {\cal F}_{Bint}^2 \rangle=0, \quad
\langle {\cal F}_{Dint}^2 \rangle=0,
\end{equation*}
if $\sigma=0$. Thus these terms do not contribute to the interaction of the inclusions in the approximation. This assertion is in agreement with the results of the works \cite{GB93,GG96,PL96,FD97,NE97,DF99,KG99,YD12,HW01,GW21}. 

However, the cross term coming from the energies (\ref{inter1}) and (\ref{energy4}) is non-zero \cite{GB93}. The corresponding contribution into the interaction energy is determined by the correlation function
\begin{eqnarray}
\langle  (\nabla^2 u_1)^2
 [\partial_x^2 u_2 \partial_y^2 u_2 -(\partial_x \partial_y u_2)^2] \rangle =
 \nonumber \\
=  2 \langle \nabla^2 u_1 \partial_x^2 u_2\rangle \langle \nabla^2 u_1 \partial_y^2 u_2\rangle
 \nonumber \\
 -2 \langle \nabla^2 u_1 \partial_x \partial_y u_2 \rangle ^2
 =-\frac{T^2}{2\pi^2 \kappa^2 r^4},
 \nonumber
 \end{eqnarray}
where we substituted the expressions (\ref{aper5}-\ref{aper10}). Thus, the pairwise interaction potential is
\begin{equation}
U_{12}= \frac{(T B_1 D_2 + T B_2 D_1)}{2\pi^2 \kappa^2 r^4}.
\label{inter22}
\end{equation}
Here and afterward $r=|\bm r_1-\bm r_2|$. This case was considered in the literature \cite{GB93,DF99} as well. The interaction (\ref{inter22}) is repulsive if $B_i D_j >0$, otherwise if $B_i$, $D_j$ have opposite signs, the interaction is attractive.

It is worth to noting that the interaction energy (\ref{energy}) related to the coupling term (\ref{inter1}) becomes non-zero, if we take into account the surface tension. The corresponding contribution to the interaction potential is
\begin{eqnarray}
U_{12}=-\frac{1}{T}B_1 B_2 \left\langle
 [\nabla^2 u(\bm r_1)]^2 [\nabla^2 u(\bm r_2)]^2 \right\rangle
 \nonumber \\
=-\frac{T B_1 B_2 \sigma^2}{2\pi^2 \kappa^4}
\left[K_0(\sqrt{\sigma/\kappa}\,r)\right]^2,
\label{inter2}
\end{eqnarray}
in accordance with Eq. (\ref{aper3}). The factor in front of the interaction potential (\ref{inter2}) is small in comparison with the factor entering  Eq. (\ref{inter22}) due to the smallness of the surface tension $\sigma$. On distances less than $\sqrt{\kappa/\sigma}$ the interaction potential (\ref{inter2}) is logarithmic. Note also that the interaction (\ref{inter2}) is attractive for the identical inclusions.

\subsection{Inclusions with quadrupolar in-plane anisotropy.}
\label{subsec:ii}

The class (ii) is partially known from the literature \cite{PL96} for the inclusions in the tensionless membranes. For completeness we outline shortly the results of the calculations. If the protein molecule cross-section is anisotropic, there appear additional interaction terms. If the shape of the inclusion cross-section possesses quadrupole symmetry the anisotropic coupling terms are proportional to the symmetric traceless second order tensor $2n_i n_k - \delta _{ik}$. Here $\bm n_j$ is the unit vector characterizing the in-plane anisotropy of the protein cross-section. There are at least two anisotropic contributions to the interaction energy
\begin{eqnarray}
{\cal F}_{Mint}= \sum_j   M_j \nabla^2 u \,(2n_{ji}n_{jk}-\delta_{ik})\partial_i\partial_k u,
\label{energy55} \\
{\cal F}_{Hint} = \, \sum _j H_j \left[ (2n_{ji}n_{jk}-\delta_{ik})\partial_i\partial_k u \right]^2,
\label{energy5}
\end{eqnarray}
in addition to the isotropic terms (\ref{inter1},\ref{energy4}).

One can find quadratic and cross contributions to the interaction potential $U_{12}$ related to the interaction energies (\ref{inter1},\ref{energy4},\ref{energy55},\ref{energy5}). The large number of various cases with different kinds of in-plane anisotropy for the both up-down symmetric and up-down asymmetric protein molecules, have been reported in the literature \cite{PL96}. For the up-down symmetric protein molecules we retrieve from the general expression (\ref{energy}) the interaction potential $\propto r^{-4}$.
A more specific studies and detail analysis of the in-plane anisotropic interactions become appropriate if suitable experimental results will become available.

If the in-plain anisotropy of the proteins is weak, then the main anisotropic contribution into the interaction potential between inclusions is linear in $M_j$ (\ref{energy55}). Then, the main contribution into the pairwise interaction energy is determined by the cross-coupling terms between the anisotropic part of the coupling energy (\ref{energy55}) and two isotropic contributions (\ref{inter1},\ref{energy4}). The average $-T^{-1}\langle {\cal F}_{Mint} {\cal F}_{Bint}\rangle$ is zero, see (\ref{oper2}), therefore the non-zero cross-coupling term arises from the average $-T^{-1}\langle {\cal F}_{Mint} {\cal F}_{Dint}\rangle$. Thus we end up with the following interaction potential
\begin{eqnarray}
U_{12}=-\frac{TM_1D_2}{\pi^2\kappa^2 r^6}
[2(\bm r \bm n_1)^2-r^2]
\nonumber \\
-\frac{TM_2D_1}{\pi^2\kappa^2 r^6}
[2(\bm r \bm n_2)^2-r^2].
\label{weak}
\end{eqnarray}
For the identical inclusions $M_1 = M_2 = M$ and $D_1 = D_2=D$, and the interaction energy (\ref{weak}) acquires the following form
\begin{eqnarray}
U_{12}=-\frac{4 T M D}{\pi^2\kappa^2 r^4}
(1 + \cos^2\phi _1 + \cos^2 \phi _2) .
\label{weak1}
\end{eqnarray}
Here $\phi _1$ is the angle between $\bm r={\bm r}_1 - {\bm r}_2$ and ${\bm n}_1$, and $\phi _2$ is the angle between ${\bm r}_1 - {\bm r}_2$ and ${\bm n}_2$.

As it follows from the expression (\ref{weak1}), if the product $M D$ is positive, the maximal attraction between weakly anisotropic inclusions occurs for the parallel to $\bm r$ or perpendicular to $\bm r$ orientations of the inclusions. In both cases the orientations of the inclusions are the same therefore the cases
are equivalent. (To see it, one can substitute $\bm n$ by the unit vector perpendicular to $\bm n$, it is a question of the definition of $\bm n$.) This fact suggests an energetic advantage for the linear (one dimensional) construction of the protein aggregates, reported in the literature \cite{SS13}.

\subsection{Linear in $u$ coupling}
\label{subsec:iii}

If the protein molecule is up-down asymmetric and isotropic in the plane, then there is the linear in the membrane curvature contribution into the interaction energy
\begin{equation}
{\cal F}_{Cint}= \sum_j   \,C_j \nabla^2 u(\bm r_j).
\label{inter3}
\end{equation}
The interaction energy (\ref{inter3}) accounts for the fact that locally the protein molecule might create
up-down asymmetry of the membrane, i.e., its local spontaneous curvature \cite{KL93,SA94,NT00,CL00}.

Using the general expression (\ref{energy}), one finds from Eq. (\ref{inter3}) the following interaction energy between two protein molecules, $1$ and $2$,
\begin{eqnarray}
U_{12}=-\frac {1}{T}C_1 C_2 \left\langle
 \nabla^2 u(\bm r_1) \nabla^2 u(\bm r_2) \right\rangle
 \nonumber \\
=\frac{C_1 C_2 \sigma}{2\pi \kappa^2}
K_0(\sqrt{\sigma/\kappa}\,r),
\label{inter4}
\end{eqnarray}
in accordance with Eq. (\ref{aper3}). Thus, for the identical proteins the interaction is repulsive and small due to the small factor $\sigma$. At distances $r<\sqrt{\kappa/\sigma}$ the interaction potential (\ref{inter4}) is logarithmic.

As in Subsection \ref{subsec:i}, we can generalize the interaction potential $U_{12}$ for the anisotropic in-plane protein molecules. For the quadrupole insertions, one finds the coupling term
\begin{eqnarray}
{\cal F}_{int}= \sum_j  \, G_j \left[ (2n_{ji}n_{jk}-\delta_{ik})\partial_i\partial_k u \right].
 \label{energy5-bis}
\end{eqnarray}
In this case there are two contributions to the interaction energy. First, there is the contribution $-(2T)^{-1}\langle {\cal F}_{Gint}^2\rangle$ and, second, there is the contribution from the coupling terms (\ref{inter3}) and (\ref{energy5-bis}). Both contributions are $\propto r^{-2}$ unlike $\propto r^{-4}$ for the up-down symmetric proteins. Let us mention that, unlike the potential (\ref{inter4}), these interaction potentials do not have a smallness in $\sigma$.

If the anisotropy is weak, then the main anisotropic contribution to the interaction potential is determined by the cross term $-T^{-1} \langle {\cal F}_{Cint} {\cal F}_{Gint}\rangle$. It gives the interaction potential
\begin{equation}
U_{12}= - \frac{1}{\pi \kappa r^2}
\left(\delta_{km}-2\frac{r_k r_m}{r^2}\right)
(G_1 C_2 n_{1k}n_{1m}+G_2C_1 n_{2k}n_{2m}),
\label{aniso}
\end{equation}
where, as above $\bm r=\bm r_1-\bm r_2$. For identical insertions, the expression (\ref{aniso}) leads to the angular factor
\begin{eqnarray}
U_{12} \propto\cos (2 \phi _1) + \cos (2 \phi _2),
\label{new17}
\end{eqnarray}
where, as above, the angles $\phi_1$ and $\phi_2$ are the angles between $\bm n_1$ and $\bm r$ and  between $\bm n_2$ and $\bm r$, respectively. The results are in accordance with Refs. \cite{PL96,DF99,DF02}. Similar to the up-down symmetric case considered in the previous section, the attraction potential for the identical anisotropic up-down asymmetric inclusions achieves its maximal value for $\phi _1 = \phi _2 = 0$, or $\phi _1 = \phi _2 = \pi /2$, i.e., there is a tendency to the identical orientation of the insertions and, consequently, to their linear aggregation.

As a note of caution we would like to add that due to coupling term (\ref{inter3}) a spontaneous curvature of the membrane has to be generated in the dense domains of the inclusions.  This effect certainly plays an important role for protein aggregation processes in living cells.
A number of striking examples of electron micrograph of a cell, presented in the works
\cite{PN05,SC12} shows that the cell membranes often tend to become strongly curved. Experimental and numeric observations manifest that in such a situation one has to study also feedback effects, describing how the inclusions themselves influence membrane equilibrium shapes \cite{SS13,PN05,FG15,WB09,OS19,SC12,BS22}.
Just this spontaneous curvature, created by the proteins, is responsible for the protein distillation
in living cells: clusterization of the proteins gives rise the finite spontaneous curvature and as a result to a formation of vesicles saturated by the specific protein molecules \cite{ZV21}.
However at low densities of proteins our approximation of the nearly flat membrane looks reasonable, because in such conditions the induced spontaneous curvature is also small.

\section{Electric field induced interactions}
\label{sec:electric}

As we already mentioned, the interaction between proteins inserted in a lipid membrane can be tuned by various external fields
coupled to the membrane mechanical degrees of freedoms (area stretching, thickness compression,
shear deformation, chain tilting, and, notably, curvature deformation). In this Section we show that it is possible to change
the interaction potential between proteins by an external uniform electric field (both constant or alternating in time) provided
the protein molecule possesses an electric dipole moment ${\bm d}$. Unfortunately applying external electric field is not a
completely harmless action. First of all lipid membranes, both model and living ones,
have also anisotropic electric polarizability. Thus an external electric field can suppress the membrane shape
fluctuations, and hence, the interactions mediated by these fluctuations. Typically,
the dielectric anisotropy of lipid membranes is small, and calculating the membrane shape
fluctuations we can neglect the contribution quadratic over the external field ${\bm E}$.
However protein molecules often have their own electric dipole moments ${\bm d}$.
This dipole coupling, linear in the external electric field, could be much larger than the dielectric coupling quadratic over field.

The corresponding interaction energy is $-E d \cos \theta$, where $\theta$ is the angle between the electric field and the dipole moment. Below we assume that the dipole moment $\bm d$ is perpendicular to the membrane and that the electric field $\bm E$ is directed along $Z$-axis. Then $\cos\theta\approx 1-(\nabla u)^2/2$. Therefore the coupling energy contains the following term, linear in the external electric field and quadratic in the tilt $\nabla u$,
\begin{equation}
{\cal F}_{int}=\sum_j   \, A_j [\nabla u(\bm r_j)]^2,
\label{aver4}
\end{equation}
where the coefficients $A_j$ are proportional to the electric field and to the dipole moments, $A_j = (1/2)Ed_j$. To avoid a confusion, it is worth to note that the external electric field breaks the rotational invariance of the membrane. That is why the term (\ref{aver4}) is changed, say, under rotation around the axis $X$ by the an angle $\varphi $, when $u \to u + \varphi x$).

The interaction energy between two proteins located at the positions, $1$ and $2$, induced by the coupling term (\ref{aver4}), is
\begin{equation}
U_{12}=-\frac{1}{T}A_1 A_2 \left\langle
 [\nabla u(\bm r_1)]^2 [\nabla u(\bm r_2)]^2 \right\rangle,
\label{energy2}
\end{equation}
in accordance with Eq. (\ref{energy}). Here we assumed that the external electric field is homogeneous in space. Therefore
\begin{equation}
U_{12}=-\frac{T A_1 A_2}{(2\pi \kappa)^2}
\left[\ln\frac{1}{kr}\right]^2,
\label{energy3}
\end{equation}
in accordance with Eq. (\ref{aper2}). The cut-off wave vector $k$ entering (\ref{energy3}) is $k=(\sigma /\kappa )^{1/2}$, the expression (\ref{energy3}) is correct provided $kr$ is small. Thus we end up with the interaction potential, proportional to the squared logarithm of $r$.

If the dipole moments of the interacting proteins are parallel ${\bm d}_1 \parallel {\bm d}_2$, the coefficients $A_1$ and $A_2$ have the same signs, and the interaction energy (\ref{energy3}) is attractive. For the antiparallel dipole moments the interaction is repulsive. Thus, in the external electric field there is a tendency to segregate proteins with different directions of the dipole moments (proteins with parallel dipole moments attract each other, whereas proteins with antiparallel dipole moments repel each other). If a cluster of proteins with the identical dipole directions is formed then a spontaneous curvature appears in this region. That could lead to out-pouching the membrane and then to producing a mini-vesicle.

Note that real liquids containing lipid membranes usually have an appreciable conductivity so d.c. electric fields will be screened. To avoid the difficulty it is better to use an alternating (a.c.)
external electric field. Generalizing our results for the case of alternating electric field we have to take into account that dynamically the membrane is not a strictly two dimensional fluid. Its dynamic at distances, larger than the membrane thickness, is determined by the fluids outside the membrane (see the very influential classical work \cite{SD75}). At such scales shape fluctuations of the membrane are described by the overdamped mode with the following dispersion law \cite{KL93,SA94}
\begin{equation}
\omega \sim i {\kappa q^3}/{\eta },
\label{b}
\end{equation}
where $q$ is the wave vector and $\eta$ is the dynamical viscosity coefficient of the surrounding liquid. 

The a.c. external field modifies the character of the membrane fluctuations mediating the interaction of the inclusions. However, the modification is irrelevant at scales $r$, satisfying $rk_\omega<1$ where
\begin{equation*}
k_\omega \left(\eta \omega /\kappa\right)^{1/3},
\end{equation*}
where $\omega$ is the frequency of the external field. The expression for $k_\omega$ is obtained from Eq. (\ref{b}). Since the energy (\ref{energy3}) is quadratic in the external electric field then the effective interaction between the proteins on times larger than $\omega^{-1}$ has the same form (\ref{energy3}). The only modification is that the product $A_1A_2$ should be substituted by one half of the product of the amplitudes of $A_1$, $A_2$. If the cut-off wavevector $k\sim (\sigma /\kappa )^{1/2}$ is less than $k_\omega$, then $k$ should be substituted by $k_\omega$ in the argument of the logarithm in (\ref{energy3}).

Our claim in this section and the main new message of our work is that the external electric field (both d.c. or a.c.)
can enhance considerably the interaction, mediated by the membrane shape fluctuations,  as the expression (\ref{energy3}) represents
strong long-range interaction without small parameter $\sigma$. We hope that there are some bridges of our results to existing knowledge in the vast field of science studying aggregation phenomena in dilute systems. To this point it is worth to mention the very recent paper on electric field induced macroscopic two-dimensional cellular phase separation in a suspension of nanoparticles with very low volume fraction \cite{RC22}. The key point of that work (similar to our results presented in this Section) is that the interaction is strongly enhanced in the external electric field.

\section{Estimations of the coupling constants}
\label{sec:estim}

A whole wealth of information about phenomenological constants ($A\, , B\, ,C\, , D\, , G\, , H\, , M$ entering the coupling terms and the interaction potentials, presented in Sections \ref{sec:fluct} and \ref{sec:electric} can be found only from ab-initio microscopic computations or from experimental data. The both approaches are beyond our skills. Thus in this Section \ref{sec:estim} we restrict ourselves only to qualitative dimensional estimations of the coupling constants. We do believe that the estimations help to understand crudely the relevance
of the discussed coupling mechanisms (i)-(iv), and to avoid fallacy in their interpretation.

Having in mind inclusions like freely rotating up-down symmetric proteins, the main mechanism of the coupling between protein and membrane,
is so-called hydrophobic mismatch the papers \cite{Mouritsen84,Gordeliy}. This mismatch produces the local membrane thickness variation
\begin{equation}
h = h_0(1+\epsilon),
\label{m1}
\end{equation}
where $h_0$ is the equilibrium bare membrane thickness, and $\epsilon $ (assumed small in our perturbation approach) is
dimensional parameter which characterizes the hydrophobic mismatch.
In own turn the membrane thickness variation changes locally the membrane rigidity
moduli $\kappa$ and ${\bar \kappa }$. According to the elasticity theory of thin shells \cite{LL84}  $\delta \kappa \sim \kappa  \epsilon$.
Therefore in the limit $\epsilon \ll 1$, we expect the coupling parameters $B_j \sim \kappa  h_0^2 \epsilon $, thus positive or negative depending on the sign of the mismatch. The presented above scaling for $B_j$ is based on the natural dimensional estimates in terms of the Helftich modulus $\kappa$ and the membrane thickness $h_0$.

Similarly by dimensional arguments we estimate the mean-Gaussian rigidity cross-coupling coefficient $D_j$. Very roughly, the part of a protein molecule
that is embedded within the bilayer, can be modeled \cite{FG15} by a disc, and the membrane contact angle at the disc
is determined by the tendency of the membrane lipid molecules to align on the protein molecule. Assuming again that
the main role in the coupling is played by the hydrophobic mismatching $\epsilon$ we estimate  coefficients $D_j$ as
$D_j \sim \kappa h_0^2 \epsilon $. Note that the sign of the coupling constant is determined by the sign of the $\epsilon $.

The coefficients $H,\, M$ can be estimated as $\kappa h_0^2$ times the relative anisotropy, which can be characterized for the inclusions having quadrupolar in-plane symmetry
by a single scalar dimensionless parameter $s$. Therefore the dimensional estimation for
these coefficients reads as $H,\, M \sim \kappa h_0^2 s$. The natural dimensional estimations for the coefficients $C$ and $G$ should include
besides the factor $\kappa h_0$ also the value of the cross-section anisotropy (i.e., $s$) as well as new parameter
$\delta $, which characterizes the relative breaking the up-down symmetry. Namely $C \sim \kappa h_0 \delta $ and $G \sim \kappa h_0 s \delta $.
Note that these estimates are proportional to the first power of $h_0$ unlike to the cross-coupling (bending rigidity - Gaussian rigidity) term.

Finally, the coupling coefficient $A$, induced by electric field, scales as $d E_0$ where $d$ is the dipole moment and $E_0$ is the external electric field.
Unfortunately, these qualitative and pure dimensional estimations are not able to catch possible small factors (both numeric and due
to existence of several dimensionless parameters, like $h_0/a$,  where $a$ is the thickness of the inclusion.

\section{Conclusion}
\label{sec:con}

In our work we revisited a phenomenological theory of membrane mediated interaction between well separated inclusions. We reproduced some known results \cite{GB93,GG96,PL96,FD97,NE97,DF99,KG99,YD12,HW01,GW21} derived here by pedagogically more transparent method.
Our motivation for presenting this discussion is one new result emanated from our study. Namely, the question we are interested in this work
is how it is possible to change the interaction potential by external actions. Surprisingly, this question does not seem
to have been addressed so far. As an example of such external action we consider a uniform electric field (constant, d.c., or alternating,
a.c.) coupled to the protein molecule dipole moment $\bm d$. We found that in the external electric field, membrane fluctuations
induce strong and very long range interaction potential proportional to the logarithm squared of the distance between inclusions $r$. This interaction can be attractive or repulsive for parallel or anti-parallel to the membrane normal orientation of the inclusion dipole moments, thus triggering either inclusion aggregation or segregation.

The presented theory describing pairwise interaction potential between inclusions in a membrane, has certain value in its own right.
However, we do believe that the main interest of the results is in the case when the inclusions are various trans-membrane
proteins. Such inclusions can be formed by a single protein molecule, or by a small cluster of proteins (a few molecules).
Very often the membrane is not homogeneous itself. It contains regions with excess concentrations of lipids (so-called rafts \cite{NT00}). Membrane elastic moduli within these regions can be different from those of the surrounding membrane. For this reason membrane fluctuations, affected by such inclusions, induce elastic interactions between the rafts.
For living cells our approximation $r \gg a$ (where $a$ is the thickness of the inclusion) in a rigorous meaning does not work for typical size of the rafts $a \simeq  100\, nm$.
The obtained pair-wise potentials can be used only to catch some qualitative features of the interactions. However, the condition
$r \gg a$ can be satisfied for artificial lipid membranes used in realistic experimental conditions.

It is worth to note that protein-induced lipid bilayer thickness deformations (hydrophobic mismatch)
provide a general physical mechanism coupling lipid and protein organization in bilayers with heterogeneous hydrophobic
thickness. The mehanism yields, without any assumptions about preferential interactions between particular lipid and protein species,
to organization of lipids and membrane proteins according to their preferred hydrophobic thickness \cite{EN05,MK20,SK20}. Combining
hydrophobic mismatch coupling and electric field induced protein dipole moment coupling leads to a possibility
to handle simultaneously membrane lipids and trans-membrane proteins organization.

As it is well known, the membrane proteins and their aggregates are key ingredients of almost all biological functions. It is sufficient
to mention (the topic especially fashionable nowadays) that just specific protein clusters support immune protection).
Although active molecular motors play an important role in the aggregation processes within biological cell, at least at the initial step of
the aggregation the interactions between protein molecules may provide a powerful aggregation mechanism.
Unfortunately, the components of the real biological membranes and real proteins are too diverse and complex to obtain detailed and unambiguous
information about protein aggregation phenomena. Thus, complementary simple theoretical approaches, describing model (single lipid) membrane systems
are necessary to elucidate the role of the lipid bilayer in the initial processes of protein aggregation.
 
There is accumulated a quite large body of theoretical works on interactions
between various objects (molecules) attached somehow or embedded into lipid membranes (see e.g., not exhaustive list of publications, partially
discussed in our paper, \cite{GB93,GG96,PL96,FD97,NE97,DF99,KG99,YD12,HW01,GW21}).
Our simple theory predicts some measurable signatures of the inter-protein interactions, e.g., their dependencies
on the geometrical parameters of the protein molecule (its radius $a$, up-down asymmetry $\delta$, in-plane
anisotropy $s$, or its electric dipole moment $d$), membrane elastic moduli $\kappa $ and $\bar {\kappa }$, and equilibrium thickness $h_0$, as well as
(as we do believe convinced by the results presented in the work \cite{Gordeliy}) the main controlling interaction parameter is the hydrophobic mismatch $\epsilon $. The macroscopic theory does not allow to calculate the phenomenological coupling coefficients
($A\, ,\, B\, , C\, ,\, D\, ,\, G\, ,\, H\, ,\, M$). These coefficients are determined by the short-range contributions and local structure of the lipid
layer around the protein molecules \cite{WB09}. Some qualitative estimations of these coefficients are presented in the previous section
\ref{sec:estim}.

Even in the framework of our phenomenological approach a number of questions remains to be clarified.
For example, protein molecules in a liquid membrane can freely rotate. Besides,
proteins may be tilted in the membrane, and their orientation may fluctuate. The former effect averages out the in-plane anisotropic contributions,
however, in the second order over anisotropic part of the interaction energy, it creates angular correlations between the interacting
proteins. In own turn these correlations can facilitate considerably the protein crystallization \cite{OS19,SC12,SS13,BS22}. The dipole moments of the 
protein molecules are subject of further investigations. The matter is that the ferroelectricity has long been speculated to have important
biological functions see e.g., \cite{LZ13}. Found in our work electric field induced giant amplification of the pairwise potential
between bearing dipole moment inclusions (\ref{energy3}), suggests a principle possibility to achieve a ferroelectric or an antiferroelectric
ordering of the protein molecules embedded into the membrane.

Having in mind living cells in this work we considered inclusions inserted in a single membrane.
However in experiments on protein crystallization, lamellar, cubic, or sponge phases are usually
utilized \cite{CS08,MG14,Gordeliy}. Two dimensional membrane shape fluctuations (the mechanism, providing
inter-protein interactions) for such non-uniform structures have a natural long-range cut-off.
In order for there to be an area of applicability of our approach, certain conditions must be met.
For the lamellar (smectic-like) structure it requires that the inter-membrane distance $d_0$ is much larger than the layer thickness
$h_0$. For a sponge or cubic structures $d_0$ is a characteristic distance (lattice size in the cubic phase)
between topological ``hands'' (or saddles) occurring at $\bar \kappa >0$ \cite{SA94}.

The value of $d_0$ in the cubic phase can be estimated by simple dimensional arguments. The matter is that the Helfrich energy (\ref{Helfrich}) does not change at self-similar deformations,
keeping constants the membrane mean curvature and the number of topological hands per unit cell. Therefore to
estimate the energy cost of such deformations, we have to take into consideration the next order terms over the curvature
expansion. It yields to the characteristic elastic energy $\kappa h_0^2/d_0^2$. To find the maximal possible cubic
lattice size, this energy should be compared to the entropy, which is on the order of $k_B T$ per unit cell.
It gives us the estimation of the maximal value of $d_0$ in the cubic structure
\begin{equation}
d_0^2 \simeq \frac{h_0^2 \kappa }{k_B T} ,
\label{new5}
\end{equation}
Since typically $\kappa \simeq 10^2 \, K_B T$, there is a quite broad range of parameters, where our results
are applicable for the cubic phase. All the more said above is true for a sponge phase, which can be considered
according to the Lindemann criterion (\ref{new5}) as a melted at  scales larger than $(\kappa h_0^2/k_B T)^{1/2}$ cubic structures.

More detailed investigation is beyond the scope of our work. We do believe that the described in our work mechanism of electric field enhanced inter-particle interaction can bring about many scenarios of protein aggregation worthy of further studies. The direct experimental measure of such interaction potential
is still a challenging task. Based on the evidence for mismatch induced interaction yielding to protein clustering \cite{Gordeliy},
we hope that our paper will stimulate discussions on the intriguing and important issues of membrane-proteins aggregation and crystallization.

\acknowledgements

Our understanding of mismatch phenomenon (i.e., length difference between membrane thickness and hydrophobic part of protein molecules inserted into the membrane), benefited tremendously from stimulating discussions with V.Gordeliy. The work of E.S.P., E.I.K. and V.V.L was supported by the State assignment N. 0029-2019-0003, the work of A.R.M. was supported by the State assignment  FMME-2022-0008  (N. 122022800364-6).

\appendix

\section{}
\label{sec:pair}

Starting from the expression (\ref{energy1}) one obtains that the pair correlation function of $u$ in Fourier representation is $T/(\kappa q^4 +\sigma q^2)$. Performing the inverse Fourier transform, one finds
\begin{eqnarray}
\langle \nabla u(\bm r_1) \nabla u(\bm r_2) \rangle
=\int \frac{d^2 q}{(2\pi)^2}
\frac{T q^2}{\kappa q^4 +\sigma q^2}
e^{i \bm q  \bm r}
\nonumber \\
=\frac{T}{2\pi\kappa}
K_0(kr),
\label{aper1}
\end{eqnarray}
where $\bm r=\bm r_1-\bm r_2$, $k=\sqrt{\sigma/\kappa}$. Next, one finds
\begin{eqnarray}
\langle \nabla^2 u(\bm r_1) \nabla^2 u(\bm r_2) \rangle
=\int \frac{d^2 q}{(2\pi)^2}
\frac{T q^4}{\kappa q^4 +\sigma q^2}
e^{i \bm q  \bm r}
\nonumber \\
=\frac{T}{\kappa}\delta(\bm r) -\frac{T\sigma}{2\pi\kappa^2}
K_0(kr).
\label{aper3}
\end{eqnarray}
At finite $r$ the $\delta$-function in Eq. (\ref{aper3}) can be ignored. Next, one obtains
\begin{eqnarray}
\langle \partial_i u \partial_k u \rangle
=\int \frac{d^2 q}{(2\pi)^2} \frac{Tq_i q_k}{\kappa q^4+\sigma q^2} \exp(i\bm q \bm r)
\nonumber \\
=\frac{T}{2\pi\kappa}\left\{\left[\frac{1}{k^2r^2}-\frac{1}{kr}K_1(kr)\right]\delta_{ik}
-\left[\frac{2}{k^2 r^2}-K_2(kr)\right]\frac{r_ir_k}{r^2}\right\}
\nonumber \\
\approx \frac{T}{4\pi \kappa}\left(\delta_{ik} \ln\frac{1}{kr}-\frac{r_i r_k}{r^2}\right),
\label{aper2}
\end{eqnarray}
where we kept principal terms in $kr$, assuming $kr\ll1$.

We introduce
\begin{eqnarray}
S_{in,km}(\bm r)=
\langle \partial_i\partial_n u(\bm r_1) \partial_k \partial_m u(\bm r_2) \rangle
\label{oper} \\
=T\int \frac{d^2q}{(2\pi)^2}\frac{q_i q_kq_nq_m}{\kappa q^4+\sigma q^2}
\exp(i\bm q \bm r).
\nonumber
\end{eqnarray}
In the limit $kr\ll 1$ we find from Eq. (\ref{aper2})
\begin{eqnarray}
S_{in,km}(\bm r)
=\frac{T}{4\pi \kappa r^2}\left[\delta_{ik}\delta_{mn}
+\delta_{km}\delta_{in}+\delta_{im}\delta_{kn}
-2\delta_{ik}\frac{r_n r_m}{r^2}
-2\delta_{im}\frac{r_n r_k}{r^2} \right.
\nonumber \\ \left.
-2\delta_{in}\frac{r_k r_m}{r^2}
-2\delta_{km}\frac{r_n r_i}{r^2}
-2\delta_{kn}\frac{r_i r_m}{r^2}
-2\delta_{nm}\frac{r_i r_k}{r^2}
+8\frac{r_i r_k r_m r_n}{r^4}\right],
\label{oper1} \\
S_{ii,km}(\bm r)=S_{ik,im}(\bm r)=\frac{T}{2\pi \kappa r^2}
\left(\delta_{km}-2\frac{r_k r_m}{r^2}\right).
\label{oper2}
\end{eqnarray}
Note that there are no logarithmic terms in Eqs. (\ref{oper1}-\ref{oper2}) and $S_{ii,kk}=0$. In components
\begin{eqnarray}
\langle \partial_x^2 u \partial_x^2 u\rangle
=\frac{T}{4\pi\kappa r^6}(-x^4-6 x^2 y^2 +3 y^4),
\label{aper5} \\
\langle \partial_y^2 u \partial_y^2 u\rangle
=\frac{T}{4\pi\kappa r^6}(3x^4-6x^2 y^2 -y^4 ),
\label{aper6} \\
\langle \partial_x \partial_y u  \partial_x \partial_y u \rangle=
\frac{T}{4\pi\kappa r^6}(-x^4+6 x^2 y^2 -y^4),
\label{aper7} \\
\langle \partial_x^2 u \partial_y^2 u\rangle=
\frac{T}{4\pi\kappa r^6}(-x^4+6 x^2 y^2 -y^4),
\label{aper8} \\
\langle \partial_x^2 u  \partial_x \partial_y u \rangle
=\frac{T}{4\pi\kappa r^6}(2x^3 y -6 x y^3),
\label{aper9} \\
\langle \partial_y^2 u  \partial_x \partial_y u \rangle
=\frac{T}{4\pi\kappa r^6}(-6 x^3 y +2 x y^3).
\label{aper10}
\end{eqnarray}

\end{document}